\documentclass[a4paper,twocolumn,showkeys,floatfix,aps,prl,reprint,groupedaddress]{revtex4-1}
\usepackage{amsmath,amssymb}
\usepackage{graphics,graphicx}
\usepackage{dcolumn,bm}
\usepackage{psfrag}
\usepackage{enumerate}

\begin{document}

\title{Failure time in heterogeneous systems}

\author{Subhadeep Roy${}^{1}$}
\email{sroy@imsc.res.in}
\author{Soumyajyoti Biswas${}^{2}$}
\email{soumyajyoti.biswas@ds.mpg.de}
\author{Purusattam Ray${}^1$}
\email{ray@imsc.res.in}
\affiliation{
${}^1$ The Institute of Mathematical Sciences, Taramani, Chennai-600113, India.\\
${}^2$ Max Planck Institute for Dynamics and Self-Organization, Am Fassberg 17, G\"ottingen 37077, Germany.
}

\date{\today}

\begin{abstract}
We show that the failure time $\tau_f$ in fiber bundle model, taken as a prototype of heterogeneous materials, depends crucially on the strength of the disorder $\delta$ and the stress release range $R$ in the system. For $R$ beyond a critical value $R_c$ the distribution of $\tau_f$ follows Weibull form. In this region, the average $\tau_f$ shows the variation $\tau_f \sim L^{\alpha}$ where $L$ is the system size. For $R<R_c$, $\tau_f\sim L/R$. We find that the crossover length scale has the scaling form $R_c \sim L^{1-\alpha}$. This scaling has  been found to be valid for various disorder distributions. For $\delta<\delta_c$, $\alpha$ is an increasing function of $\delta$. For all $\delta \ge \delta_c$, $\alpha$=1/3.     
\end{abstract}


\maketitle


In amorphous and heterogeneous materials, crack growth involves complex interplay of micro-crack nucleation, growth and coalescence \cite{Freund}. In these materials the process of fracture exhibits typically the following three stages: (i) initiation and formation of the micro cracks at soft points of the sample, (ii) coalescence of the micro-cracks and (iii) the propagation of the as-formed cracks \cite{Shiyoya}. When there are no large cracks, the breaking takes place randomly throughout the body,  independent of each other, similar to what happens in a percolation process. When the micro-crack density becomes large, they coalesce and form an initial crack. At this point, the micro-cracks tend to form in the vicinity of the crack tip (determined by the stress concentration and disorder). The large crack then grows (like that in a nucleation process) and the density of micro-cracks in the sample starts decreasing as more and more micro-cracks join the large crack. Recent studies have paid lots of attention on when the fracture will be percolating type and when it will be of nucleating type \cite{Moreira,Shekhawat}. While these
steps eventually leads to fracture, each of them takes finite time
to complete, the estimates of which are crucial regarding the stability and safety
of a disordered sample. 

The time to fracture, $\tau_f$, at a particular load,  is an outcome of this spatial and temporal micro cracking dynamics mentioned above. It is defined to be the time taken for the system to fracture under a certain loading condition. It is very important from the point of view of understanding failure of a specimen and engineering design and reliability \cite{Pauchard,Bonn,Lucantonio,Curtin1,Curtin2,Guarino}. Many models have been proposed to predict the failure time \cite{Gobulovic,Pomean,Rabinovitch,Curtin2,Gurino1,Politi,Taylor}. However most models and studies are concerned with the dependence of $\tau_f$ on applied load or on the macroscopic parameters like temperature, pressure, considering the crack growth as an activation process \cite{Guarino,Rabinovitch}. The understanding of the failure time from the microscopic fracture dynamics, specially in heterogeneous materials, has remained unclear \cite{Dieter,Lawn}.

Time to failure of a mechanical system consisting of parallel members have been studied in the past \cite{Phoenix1,Bazant}. These studies mostly has the order statistics of the failure time distribution for a single member. Damage evolution and time to failure have been investigated in a model where the damage formation is a stochastic event with the probability of failure at a point $i$ at time $t$ is proportional to $\sigma_i^\eta(t)$, where $\sigma_i(t)$ is the local stress at $i$ at time $t$ \cite{Curtin2}. The model predicts two regimes of failure : percolation like failure for $\eta \le 2$ and failure with precursory 
avalanches for $\eta>2$. A numerical study of the model on a two-dimensional triangular spring network model shows that for $\eta \le 2$, the failure time is independent of the system size $L$, whereas for $\eta>2$, the failure time scales as $(ln L)^{1-(\eta/2)}$ \cite{Curtin1}. The various modes of failure: percolation, nucleation, avalanche, catastrophic or quasi-brittle like at various disorder strengths and stress redistribution ranges have been discussed in the context of failure in heterogeneous systems \cite{Moreira,Shekhawat,Sroy,Biswas}. To get understanding of the failure time of heterogeneous materials, we need to know how failure time behaves for various modes of fracture.    

In this Letter we ask the question : how failure time of a system depends on the basic microscopic parameters like the strength $\delta$ of disorder and the stress-redistribution range $R$ in the system. While these two major ingredients are known to be determining the failure modes of disordered solids \cite{Sroy,Biswas}, their effects on the failure time is less known. To this effect we have studied a simple model for fracture, namely the fiber bundle model, which has served as a generic model in reproducing many features of fracture observed in experiments on heterogeneous materials \cite{Book}. We arrive at
general form of scaling for the average failure time and also show the extreme-statistics nature of their distributions.



Fiber bundle model \cite{RMP,Pierce} consists of fibers  between two parallel bars. One bar is kept fixed while the other one is pulled with an external stress $\sigma$. Disorder is introduced in the model as the fluctuation in the strength of individual fibers. When the applied stress crosses the threshold strength of a fiber, that fiber breaks irreversibly. The stress of that broken fiber is then redistributed among the remaining intact fibers. There are mainly two ways the stress redistribution has been studied in the past: equal load sharing scheme (ELS) and local load sharing scheme (LLS). In ELS scheme the stress is redistributed among all other surviving fibers in equal amount \cite{Pierce,Daniels}. In the LLS scheme only the nearest surviving neighboring fibers of the broken one carries the extra load \cite{Phoenix,Smith,Newman,Harlow2,Harlow3,Smith2}. After such redistribution there might be further breaking of fibers otherwise, the applied stress is increased to break the next weakest fiber and the process continues until all fibers break. 

\begin{figure}[ht]
\centering
\includegraphics[width=7cm, keepaspectratio]{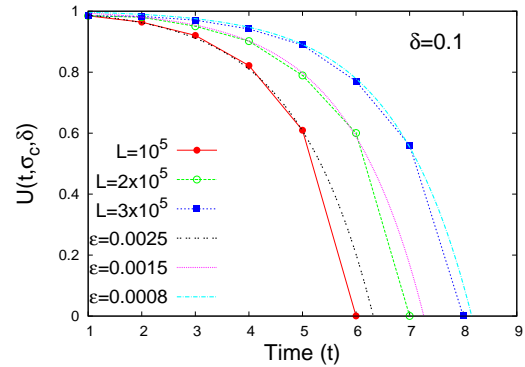} 
\caption{(Color online) Study of $U(t,\sigma_c,\delta)$ with increasing time steps for system sizes $10^5$, $2\times10^5$ and $3\times10^5$. Disorder is kept fixed at $\delta=0.1$. The black dotted lines show the analytical behavior according to Eq. \ref{eq8} with $\epsilon=0.0025$, $\epsilon=0.0015$ and $\epsilon=0.0008$ respectively. The envelop of the curve increases as well as $\epsilon$ decreases as we go to higher system sizes.}
\label{Fraction_Unbroken}
\end{figure}

The effect of $\delta$ on the time evolution of $U(t,\sigma,\delta)$, the fraction of surviving fibers at time $t$ at the stress $\sigma$ and for the width $\delta$ of the threshold distribution, can be determined analytically. If we consider the threshold distribution to be a uniform one with half width $\delta$ then, $U(t,\sigma,\delta)$ satisfies the following recursion relation
\begin{align}\label{eq1}
U(t+1,\sigma,\delta)=\displaystyle\frac{1}{2\delta}\left((c+\delta)-\displaystyle\frac{\sigma}{U(t,\sigma,\delta)}\right)
\end{align}
where $c$ is the mean of threshold stress distribution. At critical point $\sigma$ will be replaced by the critical stress $\sigma_c$. In quasi-brittle region ($\delta\ge\delta_c$) we already have an expression for $\sigma_c$ given below
\begin{align}\label{eq2}
\sigma_c=\displaystyle\frac{\delta}{2}\left(1+\displaystyle\frac{c-\delta}{2\delta}\right)^2
\end{align}
Then at critical point in the quasi-brittle regime we get from equation \ref{eq1}  
\begin{align}\label{eq3}
U(t,\sigma_c,\delta)-U_c=\left(\displaystyle\frac{1}{2}+\displaystyle\frac{1}{4\delta}\right)t^{-1}
\end{align}
where $U_c$ is the fraction of unbroken bonds at critical stress. At $\delta=0.5$, $U_c=0.5$ and the above result matches with the behavior $U(t,\sigma_c,\delta=0.5) \sim 1/t$ observed earlier \cite{Pradhan}. Also this behavior is independent of $\delta$, as long as $\delta\ge\delta_c$. In brittle region ($\delta<\delta_c$) the picture is quite different. Due to abrupt failure in this region we have $\sigma_c=\sigma_l+\epsilon$, where $\sigma_l$ is the minimum of the distribution. $\epsilon$ is the term that takes care of the system size effect in the bundle. As we go to higher system size $\epsilon$ value decreases as the threshold of the weakest link comes closer to $\sigma_l$. The recursion relation in this case takes the form
\begin{align}\label{eq4}
U(t+1,\sigma_c,\delta)=1+A\left(1-\displaystyle\frac{1}{U(t,\sigma_c,\delta)}\right)-\displaystyle\frac{\epsilon}{2\delta U(t,\sigma_c,\delta)}
\end{align}
where $A=\displaystyle\frac{\sigma_l}{2\delta}$. As $U(0,\sigma_c,\delta)=1$ its easy to see that $U(1,\sigma_c,\delta)=\left(1-\displaystyle\frac{\epsilon}{2\delta}\right)$. Repeating this recursively, we get
\begin{align}\label{eq7}
U(t,\sigma_c,\delta)=1-\displaystyle\frac{\epsilon}{(2\delta-\epsilon)}\displaystyle\frac{1-A^t}{1-A}
\end{align}
In above expression the higher order of $\epsilon$ is neglected. Using the expression of $A$ we get 
\begin{align}\label{eq8}
U(t,\sigma_c,\delta)=1-\displaystyle\frac{4\delta\epsilon}{(2\delta-\epsilon)(6\delta-1)}\left[1-\left(\displaystyle\frac{(1-2\delta)}{4\delta}\right)^t\right]
\end{align}
In figure \ref{Fraction_Unbroken} we have compared the analytical expression of $U(t,\sigma_c,\delta)$ with the numerical behavior. As higher order terms are neglected, the analytical result does not tally with the numerical findings at large times.

Numerical results are generated for system sizes ranging from $10^3$ to $10^5$ with $10^4$ configurations. The envelop of the curve in figure \ref{Fraction_Unbroken} increases as we go to higher system size suggesting an increment in failure time when system size is increased. The analytical result (Eq.\ref{eq8}) is fitted with dotted lines for different system sizes with individual $\epsilon$ values.    


The failure time $\tau_f$ is the envelop of the $U(t,\sigma_c,\delta)$ vs $t$ curve. We have determined it numerically for different $\delta$ and $R$. Numerically $\tau_f$ is estimated as the number of redistributing steps through which the bundle evolves before global failure when a critical stress is applied on it. In this paper, the results are shown for uniform distribution of half width $\delta$ and mean $0.5$ to assign individual thresholds of the fibers. We have chosen $R$ as the number of surviving fibers on both side of the broken one among which the extra stress of the broken fiber is redistributed \cite{Biswas}.

\begin{figure}[t]
\centering
\includegraphics[width=8.5cm, keepaspectratio]{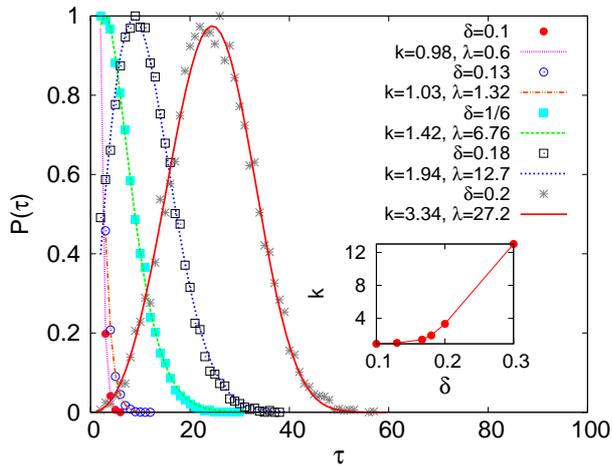} 
\caption{(Color online) The distribution for failure time for different $\delta$ is fitted with Weibull distribution with shape parameter $k$ and scale parameter $\lambda$. The numerical results are fitted with various $k$ and $\lambda$ values. In the inset the variation of shape parameter $k$ is shown with $\delta$.}
\label{Distribution}
\end{figure}

Figure \ref{Distribution} shows the distribution $P(\tau)$ of failure times $\tau$ at different $\delta$ values for system size $L=10^4$. $P(\tau)$ follows the Weibull distribution: $P(\tau) = \left(\displaystyle\frac{k}{\lambda}\right)\left(\displaystyle\frac{\tau}{\lambda}\right)^{k-1}e^{-\left(\displaystyle\frac{\tau}{\lambda}\right)^k}$, where $k$ and $\lambda$ are respectively the shape and scale parameter of the distribution. Larger the $\delta$ the distribution becomes wider. The Weibull distribution of failure times in heterogeneous materials has been discussed before \cite{Curtin1,Curtin2}. 

\begin{figure}[ht]
\centering
\includegraphics[width=8cm, keepaspectratio]{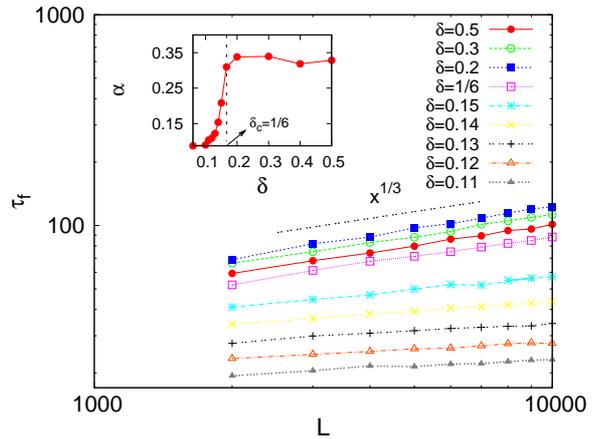} 
\caption{(Color online) System size effect of maximum average relaxation time $\tau_f$ at different $\delta$ value. $\tau_f$ shows a scale free behavior with system size ($L$) : $\tau_f \sim L^{\alpha}$. Value of exponent $\alpha$ remains constant at $1/3$ for $\delta>\delta_c$, while it keeps decreasing below $\delta_c$ and the system size effect of $\tau_f$ gradually vanishes. \\
Inset: Variation of scaling exponent $\alpha$ with varying disorder value. For $\delta>\delta_c$, $\alpha$ saturates at a value $1/3$. Below $\delta_c$ the exponent value decreases as we go to lower $\delta$ values.}
\label{Relaxation}
\end{figure}

Figure \ref{Relaxation} shows the system size effect of the average failure time ($\tau_f=\langle\tau\rangle$) at different disorder values. $\tau_f$ $\sim L^{\alpha}$ for all $\delta$ with $\alpha$ as the exponent of the power law. Above $\delta_c$, $\alpha$ shows an universal behavior and remain constant independent of $\delta$. In vanishingly small disorder the model is bound to fail in redistributing step independent of  system size. As the model approaches this vanishingly small disorder limit ($\delta \rightarrow 0$) the exponent $\alpha$ decreases. $\tau_f$ satisfies the following scaling behavior: 
\begin{equation}
\tau_f \sim L^{\alpha} 
\begin{cases}
\alpha=1/3, & \delta \ge \delta_c \\
\alpha=\Phi_{-}(\delta), & \delta < \delta_c
\end{cases}
\end{equation}
where $\Phi_{-}(\delta)$ decreases with decreasing disorder values. For uniform threshold distribution with $\delta = \frac{1}{2}$, the relaxation time at critical stress has been found before to diverge as $L^{1/3}$ \cite{chandreyee}. At 
$\delta_c = \frac{1}{6}$, the relaxation is also found to diverge as $L^{1/3}$ \cite{Sroy}. 

\begin{figure}[t]
\centering
\includegraphics[width=9cm, keepaspectratio]{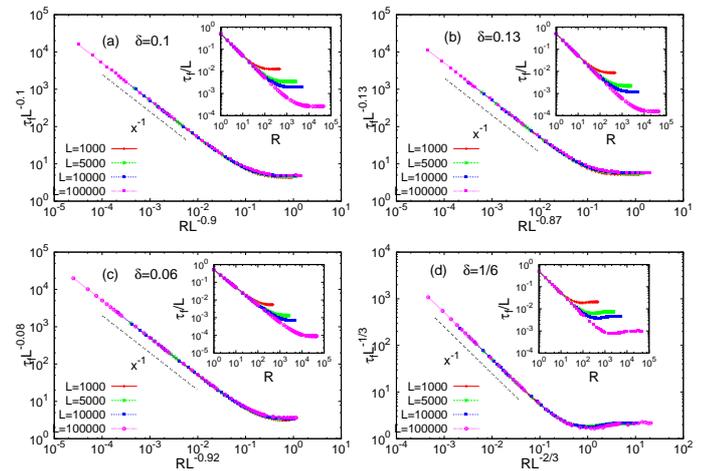} 
\caption{(Color online) Scaling of failure time ($\tau_f$) with system size $10^3$, $5\times10^3$, $10^4$ and $10^5$ for disorder (a) $\delta=0.1$, (b) $0.13$, (c) $0.06$ and (d) $1/6$. In the inset the unscaled behavior of $\tau_f$ is shown. The scaling behavior shows : $\tau_f \sim L^{\alpha}\Phi(R/L^{1-\alpha})$. Value of $\alpha$ remains the same for the region $\delta>\delta_c$. For $\delta<\delta_c$, values of this exponent $\alpha$ decreases as we go to lower $\delta$ values. In this region the scaling of $R_c$ with $L$ keep changing.}
\label{Scaling_Range}
\end{figure}

Figure \ref{Scaling_Range} shows the scaling of $\tau_f$ with range $R$ for different system sizes ranging from $10^3$ to $10^5$ for different disorder values, $\delta<\delta_c$ and $\delta>\delta_c$. We observe the following scaling of $\tau_f$ with $R$ and system size $L$ for all $\delta$ values,   
\begin{align}\label{eq:10}
\tau_f \sim L^{\alpha}f\left(\displaystyle\frac{R}{L^{1-\alpha}}\right)
\end{align} 
where $\alpha$ is the scaling exponent. The inset shows the unscaled variation of $\tau_f/L$ with $R$ at different system sizes. The scaling function $f(x)$ behaves as:
\begin{equation}
f(x) =
\begin{cases}
1/x, & R < R_c \\
Constant, & R \ge R_c
\end{cases}
\end{equation}
The value of $\alpha$ remains constant for $\delta>\delta_c$. Below $\delta_c$ this exponent becomes an increasing function of $\delta$.  

While the results presented above are for uniform threshold distribution in unit interval,   we have also checked their validity for power law and Gaussian distributions. For all these distribution the scaling relation of failure time given by Eq. \ref{eq:10} holds good. 

In conclusion, we have found  that the microscopic parameters like strength of the disorder or the range of stress redistribution has profound effect on the failure time of a disordered system. We have found the scaling relation of failure time for full parameter space of system size, strength of disorder and the stress release range in case of fiber bundle model. A universal scaling exponent is observed beyond a critical strength of the disorder value. As a result of the extreme-statistics of fracture dynamics, the distribution of failure time is found to be Weibullian over the entire parameter space.

\end{document}